\documentclass[11pt, letterpaper]{article}
\usepackage{graphicx}
\usepackage{amssymb, amsmath, amsthm}
\usepackage{algorithmic, algorithm}
\usepackage{anysize}

\begin{document}

\title{A disk-covering problem with application in optical interferometry}

\author{Trung Nguyen$^1$\thanks{The work of the author is supported by Alcatel Alenia Space and INRIA and was partly carried out while he was visiting the Freie Universit\"at Berlin.}, Jean-Daniel Boissonnat$^1$,\\Fr\'ederic Falzon$^2$ and Christian Knauer$^3$\\\noindent\\{\small $^1$Geometrica project, INRIA Sophia Antipolis, France}\\{\small $^2$Research department, Alcatel Alenia Space, France}\\{\small $^3$Institut f\"ur Informatik, Freie Universit\"at Berlin, Germany}}

\date{}
\maketitle
\marginsize{1in}{1in}{0.15in}{0.25in}

\begin{abstract}
Given a disk $O$ in the plane called the objective, we want to find $n$ small disks $P_1,\ldots,P_n$ called the pupils such that $\bigcup_{i,j=1}^nP_i\ominus P_j\supseteq O$, where $\ominus$ denotes the Minkowski difference operator, while minimizing the number of pupils, the sum of the radii or the total area of the pupils. This problem is motivated by the construction of very large telescopes from several smaller ones by so-called Optical Aperture Synthesis.  In this paper, we provide exact, approximate and heuristic solutions to several variations of the problem.
\end{abstract}

%%%%%%%%%%%%%%%%%%%%%%%%%%%%%%%%%%%%%%%%%%%%%%%%%
\section{Introduction}

\label{sec:intro}
\begin{figure}\label{fig:instr}
\begin{center}
\includegraphics[height=3cm]{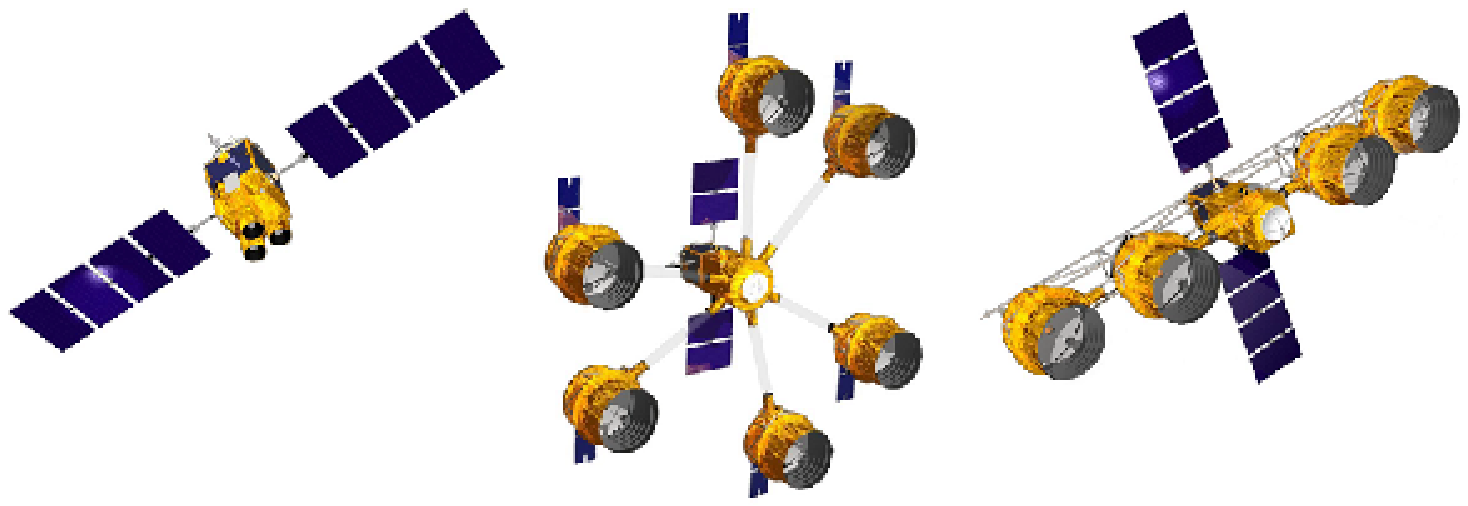}\hspace{2cm}
\includegraphics[height=3cm]{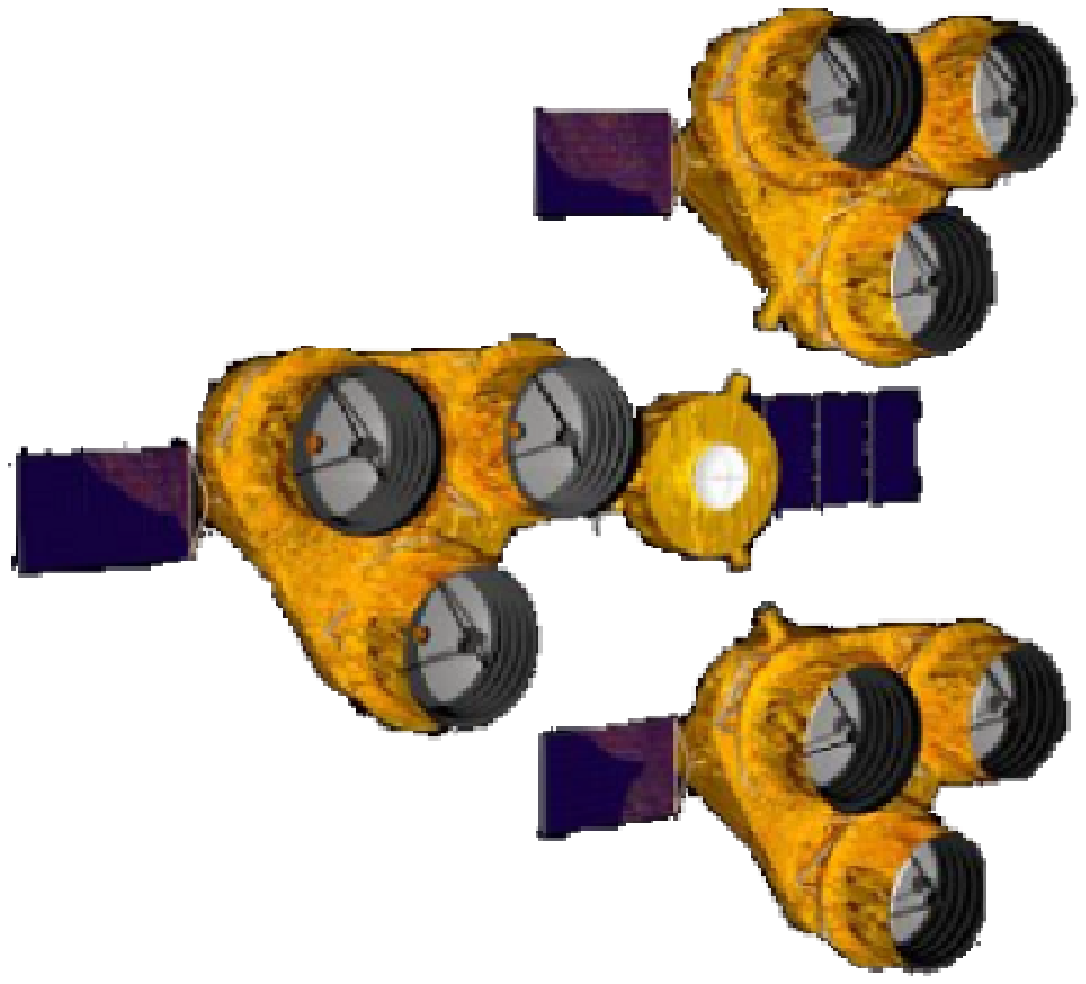}
\caption{{\small Examples of using Optical Aperture Synthesis to synthesize large pupils \cite{disrupt05}}}
\label{fig:SOO-const}
\end{center}
\end{figure}

The diameter of the pupil of a telescope is proportional to its resolution power. A simple calculus shows that we would need a telescope having a diameter of approximately $20m$ to observe the Earth from a high orbit \cite{NBBFT06}. Needless to say, such an instrument would not be adapted to the observation from space. In order not to build too large pupils, Optical Aperture Synthesis is adopted to synthesize (very) large pupils by interferometrically combining several smaller pupils \cite{disrupt05} (see Fig. \ref{fig:instr}). The auto-correlation support (ACS) of a system of pupils denotes all the observable spatial frequency domain.

The underlying problem can be stated in geometric terms as follows. Given an objective $O$ supposed to be a disk, design a set of disks $\mathcal{P}=\{P_1,\ldots,P_n\}$ such that its ACS $\mathcal{D}$ covers entirely the objective while minimizing some cost function. Here $\mathcal{D} = \bigcup_{i,j=1}^n(P_i\ominus P_j)$ where $\ominus$ denotes the Minkowski difference operator. The cost function may include the number of pupils, the sum of the radii or the total area of the pupils, etc.
This problem is a variant of the disk-covering problem. To the best of our knowledge, the variant we consider is new and the interferometry problem has not been considered before from a geometric perspective. This paper is a follow-up of our initial investigation \cite{NBBFT06}. 
The reader interested in the general disk-covering problem or some other variants can refer to \cite{alt2006mcc, CB05, booth2003cac}.

The outline of this paper is as follows. In section \ref{sec:apollonius_diagram},
we introduce Apollonius diagrams (additively weighted Voronoi diagrams) which play a central role in our study, and use them to decide whether the objective is covered.
Section \ref{sec:three_pupils} deals with the case of three pupils for which we provide an optimal solution.
We describe in section \ref{sec:approximation_number_pupils} a constant-factor approximation algorithm for the case where the pupils are restricted to have the same radius. In section \ref{sec:fixed-center_problem}, we consider the centers of the pupils to be given and provide efficient algorithms to minimize
the sum of the radii or the total area of the pupils under the constraint that the ACS covers
the objective. Finally, section \ref{sec:fixed-radius_problem} considers the problem where the radii of the pupils are known but their positions are unknown.

%%%%%%%%%%%%%%%%%%%%%%%%%%%%%%%%%%%%%%%%%%%%%%%%%%%
\section{Apollonius diagrams and the decision problem}
\label{sec:apollonius_diagram}

\newtheorem{thm}{Theorem}
\newtheorem{lem}[thm]{Lemma}
\newtheorem{cor}[thm]{Corollary}
\newtheorem{pro}[thm]{Proposition}
\newtheorem{fac}[thm]{Fact}

\subsection{Apollonius diagrams (aka Additively weighted Voronoi diagrams)}
Let $\mathcal{D}=\{ D_1,\ldots ,D_N\}$ be a set of $N$ disks in the plane.  
We denote by $c_i$ the center of $D_i$ and by $\rho_i$ its radius. Let $\|.\|$  denote the Euclidean distance and $\partial S$ denote the boundary of a subset of points $S$. The {\it distance} of a point $x$ to the circle $\partial D_i$
is defined as
\begin{displaymath}
\delta_i(x)=\|x-c_i\|-\rho_i. 
\end{displaymath}

For a point $x$, $\delta_i(x)$ is $<0,0,>0$ depending whether $x$ lies
inside, on the boundary of, or outside $D_i$. The {\it Apollonius cell} of $D_i$ consists of the points whose distance to $\partial D_i$ is less than or equal to their distance to any other circle of $\mathcal{D}$:
\begin{displaymath}
A_i=\{x\in\mathbb{R}^2\mid \delta_i(x)\leq \delta_j(x), j=1,\ldots,N \}.
\end{displaymath}

\begin{figure}
\begin{center}
\includegraphics[height=4cm]{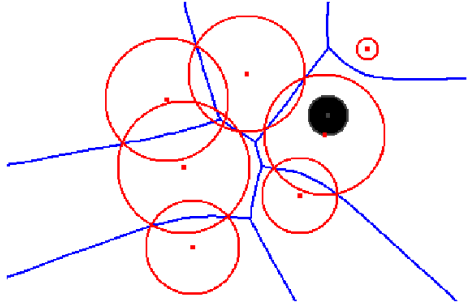}
\caption{{\small An Apollonius diagram of 8 disks in the Euclidean plane. The black disk has no cell.}}
\label{fig:power_diagram}
\end{center}
\end{figure}

Unlike the case of points, it is possible that a disk may have an empty cell. This happens when the disk is inside another disk. The one-dimensional connected sets of points that belong to exactly two Apollonius cells are called {\it Apollonius edges}, while points that belong to at least three Apollonius cells are called {\it Apollonius vertices}. The collection of the cells, edges and vertices forms the {\it Apollonius diagram} of $\mathcal{D}$, denoted by $Apo(\mathcal{D})$ (see Fig. \ref{fig:power_diagram}). The Apollonius diagram $Apo(\mathcal{D})$ can be computed in time $O(N\log N)$ which is worst-case optimal \cite{KY02}, and robust and efficient implementations exist \cite{cgal}. More information on Apollonius diagrams can be found in \cite{BWY06, KY02}. We start by stating some properties of Apollonius diagrams. Let $B_{ij}$ define the bisector of two disks $D_i$ and $D_j$
\begin{displaymath}
B_{ij}=\{x\in\mathbb{R}^2\mid \delta_i(x) = \delta_j(x) \}. 
\end{displaymath}

\begin{lem}
\label{lem:bisector_is_linear}
The restriction of $\delta_i$ and $\delta_j$ to $B_{ij}$ are unimodal functions. More precisely, these functions decrease linearly to a minimum and then increase linearly.
\end{lem}
\begin{proof}
Consider two disks $D_i$ and $D_j$ with radii $\rho_i, \rho_j$ and centers, w.l.o.g., $c_i = (-c, 0)$ and $c_j = (c, 0)$. The bisector of $D_i$ and $D_j$ is a sheet of the hyperbola whose equation is
\begin{displaymath}
\frac{x^2}{a^2} - \frac{y^2}{c^2 - a^2} = 1,
\end{displaymath}
where $a = |\rho_j - \rho_i|/2$. Then the distance of a point with abscissa $x$ on the hyperbola to $c_i$ is a linear function of $x$: $d = \pm(ex + a)$, where $e = \frac{c}{a}$ is the eccentricity of the hyperbola and sign $\pm$ is positive if $\rho_i\leq \rho_j$ and negative otherwise.
\end{proof}

\begin{cor}
\label{cor:cover_cell_by_minidisk}
Any  arc $pq$ contained in the edge of a cell $A_i$ is included in the smallest disk of center $c_i$ that contains $p$ and $q$.
\end{cor}
\begin{proof}
Since the distance function to $D_i$ of the points on arc $pq$ is unimodal by Lemma \ref{lem:bisector_is_linear}, it reaches a maximum at $p$ or $q$. Hence any disk with center $c_i$ that contains $p$ and $q$ covers the whole arc. 
\end{proof}

\begin{cor}
The Apollonius cell $A_i$ is included in the disk centered at $c_i$ that contains the set of its vertices.
\end{cor}
\begin{proof}
If $A_i$ is unbounded we are done. Otherwise, as $A_i$ is star-shaped \cite{BWY06}, it is included in a disk if its edges are. Applying Corollary \ref{cor:cover_cell_by_minidisk} to all edges of $A_i$ concludes our proof.
\end{proof}

Let $\delta_\mathcal{D}(x)$ denote the smallest distance of $x$ to the disks of $\mathcal{D}$, i.e., $\delta_\mathcal{D}(x) \leq \delta_i(x)$ for any $1\leq i\leq N$ and equality holds iff $x\in A_i$. We see that $\delta_\mathcal{D}(x)\leq 0$ when $x$ lies inside the union of the disks of $\mathcal{D}$.

\subsection{The decision problem}

\begin{figure}
\begin{center}
\includegraphics[height=4cm]{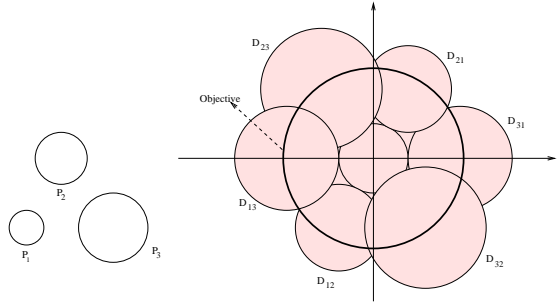}
\caption{{\small A system of three pupils (left) and its ACS (right), the objective is represented by a thick circle.}}
\label{fig:autocorrelation}
\end{center}
\end{figure}

Let $\mathcal{P}=\{P_1,\ldots,P_n\}$ be a set of $n$ disks called
the {\it pupils} and $O$ be a disk of radius $R$ centered at the origin called the {\it objective}. The ACS of $\mathcal{P}$ is $\mathcal{D}=\bigcup_{i,j=1}^n(P_i\ominus P_j)$. The decision problem consists in determining whether $O$ is covered by $\mathcal{D}$.

Let $c_i$ and $\rho_i$ denote the center and the radius of pupil $P_i$ and let  $D_{ij}=P_i\ominus P_j$. It is not difficult to see that $D_{ij}$ is a
disk with center $c_{ij}=c_i-c_j$ and radius $\rho_{ij}=\rho_i+\rho_j$. Moreover, $\mathcal{D}=\bigcup_{i,j=1}^n D_{ij}$ (see Fig. \ref{fig:autocorrelation}).

\begin{figure}[]
\begin{center}
\includegraphics[height=1.635cm]{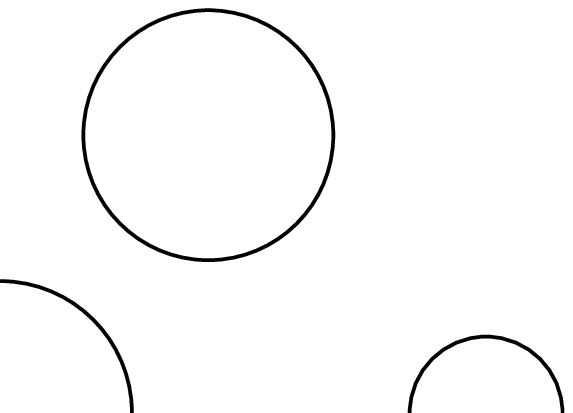}\hspace{0.25cm}\includegraphics[height=3.27cm]{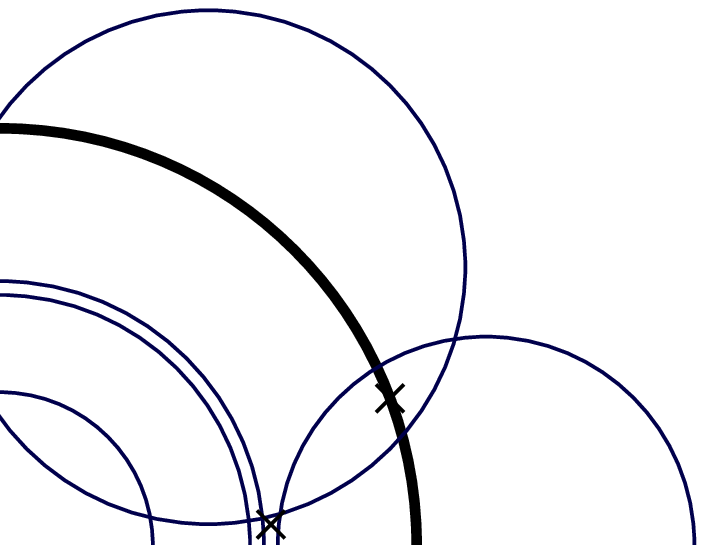}\hspace{1cm}
\includegraphics[height=1.75cm]{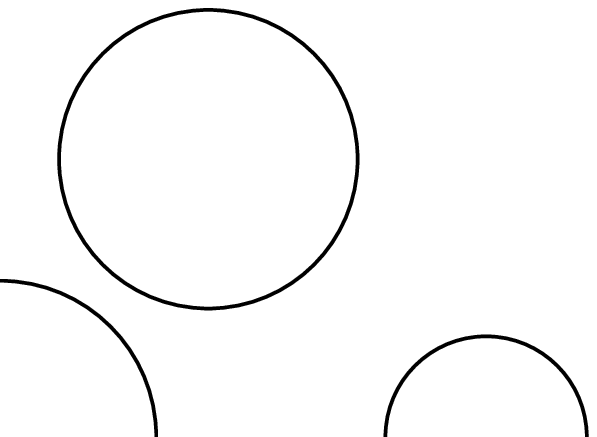}\hspace{0.25cm}\includegraphics[height=3.5cm]{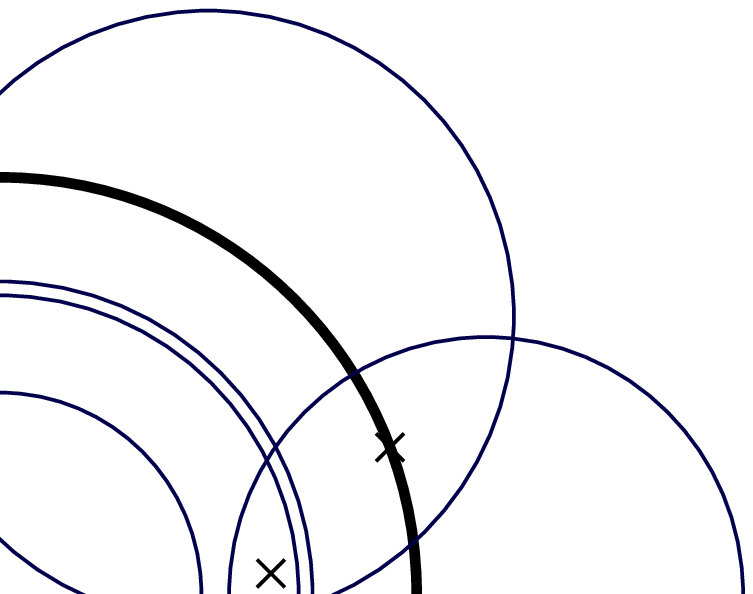}
\caption{{\small {\sc Left}: A set of three pupils whose ACS does not cover the objective. The x-marks correspond the vertices of $V_{ij}$ of which some lie outside the union of disks. {\sc Right}: The set of pupils with the same position but radii enlarged by $\alpha^*$ as computed by Algorithm \ref{alg:same_alpha}. All vertices of $V_{ij}$ are inside $\mathcal{D}$ and the objective is covered.}}
\label{fig:alpha_pupils}
\end{center}
\end{figure}

If the radius $\rho_i$ of some pupil $P_i$ is greater than half the
objective's radius $R$, $D_{ii}$ covers $O$. We assume in the sequel
that the pupils all have a radius at most $\frac{R}{2}$ which implies 
that all disks of $\mathcal{D}$ have radii  smaller than $R$. We write
$A_{ij}$ for the cell of $D_{ij}$ in the Apollonius diagram of
$\mathcal{D}$. Let $V_{ij}$ denote the set of vertices of $A_{ij}$
inside $O$ and the intersection points of $\partial A_{ij}$ with
$\partial O$. We denote by $N=n^2$ the number of the disks of
$\mathcal{D}$. It can be argued that the cardinality of all $V_{ij}$
is $O(N)$. The following shows a necessary and sufficient
condition for covering $O$ by $\bigcup_{i,j=1}^n D_{ij}$ (see
Fig. \ref{fig:alpha_pupils}).

\begin{lem}
\label{lem:decision_problem}
$O\subseteq\mathcal{D}$ iff $V_{ij}\subseteq D_{ij}$ for all $i,j=1,\ldots,n$. 
\end{lem}
\begin{proof}
First we argue that $O\subseteq\mathcal{D}$ iff $A_{ij}\cap O\subseteq
D_{ij}$ for all $i,j=1,\ldots,n$. Since the set of $A_{ij}$ forms a
decomposition of the plane, $A_{ij}\cap O\subseteq D_{ij},
i,j=1,\ldots,n$, implies that $O\subseteq
\bigcup_{i,j=1}^n D_{ij}=\mathcal{D}$. Conversely, suppose that $O\subseteq\mathcal{D}$
and $p\in A_{ij}\cap O$, we will show that $p\in D_{ij}$. Indeed,
$p\in O\subseteq \mathcal{D}$ implies $\delta_\mathcal{D}(p)\leq
0$. Together with $p\in A_{ij}$, we conclude
$\delta_{ij}(p)=\delta_\mathcal{D}(p)\leq 0$ which implies that $p$ is
inside $D_{ij}$.

We show next that $A_{ij}\cap O\subseteq D_{ij}$ is equivalent to
$V_{ij}\subseteq D_{ij}$ by proving that a disk $\Delta$ centered at $c_{ij}$
covering $V_{ij}$ covers also $A_{ij}\cap O$. We first observe that the
edges of $A_{ij}$ with both endpoints in $O$ are covered by $\Delta$ by Corollary
\ref{cor:cover_cell_by_minidisk}. It remains to verify that the
intersection points of $\partial A_{ij}$ with $\partial O$ and the
arcs linking them are also in $O$. Consider two such points $p$ and
$q$ consecutive along the boundary of $O$. Call $p_1p_2$ and $q_1q_2$
the two Apollonius edges that intersect $\partial O$ at $p$ and $q$
respectively. Suppose $p_1,q_1\in O$ and $p_2, q_2\notin
O$, which implies that $p_1, p, q, q_1$ belong to $V_{ij}$. 
Since $p_1$ and $p$ lie on edge $p_1p_2$, and $q$ and $q_1$ are contained
in $q_1q_2$, $\Delta$ will cover the arcs $p_1p$ and $qq_1$ by Corollary
\ref{cor:cover_cell_by_minidisk}. It thus remains to show that the
circular arc $pq$ of $O$ is included in $\Delta$, which is true since $p, q\in D_{ij}$
whose radius has been assumed to be smaller than the radius of $O$.
\end{proof}

%\begin{algorithm}[]
%\caption{Decide if $O$ is covered by $\mathcal{D}$}
%\label{alg:covering}
%\begin{algorithmic}[1]
%\STATE compute $Apo(\mathcal{D})$
%\IF{one of $\rho_i \geq R/2$}
%\RETURN covered
%\ELSE 
%\STATE compute $V_{ij}$
%\IF {$V_{ij}\subseteq D_{ij}$ for all $i, j=1,\ldots,n$}
%\RETURN covered
%\ELSE
%\RETURN uncovered
%\ENDIF
%\ENDIF
%\end{algorithmic}
%\end{algorithm}

The following simple result is important in sections \ref{sec:fixed-center_problem} and \ref{sec:fixed-radius_problem}.
\begin{cor}
\label{cor:disks_cover_cells}
Given a configuration of pupils with the corresponding sets $D_{ij}$ and $V_{ij}$. We move/resize the pupils such that each new disk $D^{'}_{ij}$ includes $V_{ij}$. Then, $O$ is covered by $\bigcup_{i,j=1}^n D^{'}_{ij}$.
\end{cor}
\begin{proof}
Since $V_{ij}\subseteq D^{'}_{ij}$ is equivalent to $A_{ij}\cap
O\subseteq D^{'}_{ij}$ (see the proof of Lemma
\ref{lem:decision_problem}) and the sets $A_{ij}\cap O$ cover $O$, the
objective is covered by $\bigcup_{i,j=1}^n D^{'}_{ij}$.
\end{proof}

Lemma \ref{lem:decision_problem} gives us a simple $O(N\log N)$-time algorithm that solves the decision problem. It still works when we replace Apollonius diagrams by power diagrams. The reason of using the formers will be seen in section \ref{sec:fixed-center_problem}.

%%%%%%%%%%%% SECTION %%%%%%%%%%%%%%%%%%%%
\section{Problem with three pupils}
\label{sec:three_pupils}
A configuration of pupils is called valid if its ACS covers the objective. In this section, we want to minimize the sum $\rho_1+\rho_2+\rho_3$ among the valid configurations. Let denote by $l$ the line passing through $c_{23}$ and $c_{32}$. Since the disks and the objective are symmetric about the origin, it suffices to consider only one half-plane bounded by $l$.

\begin{figure}
\begin{center}
\includegraphics[height=3.5cm]{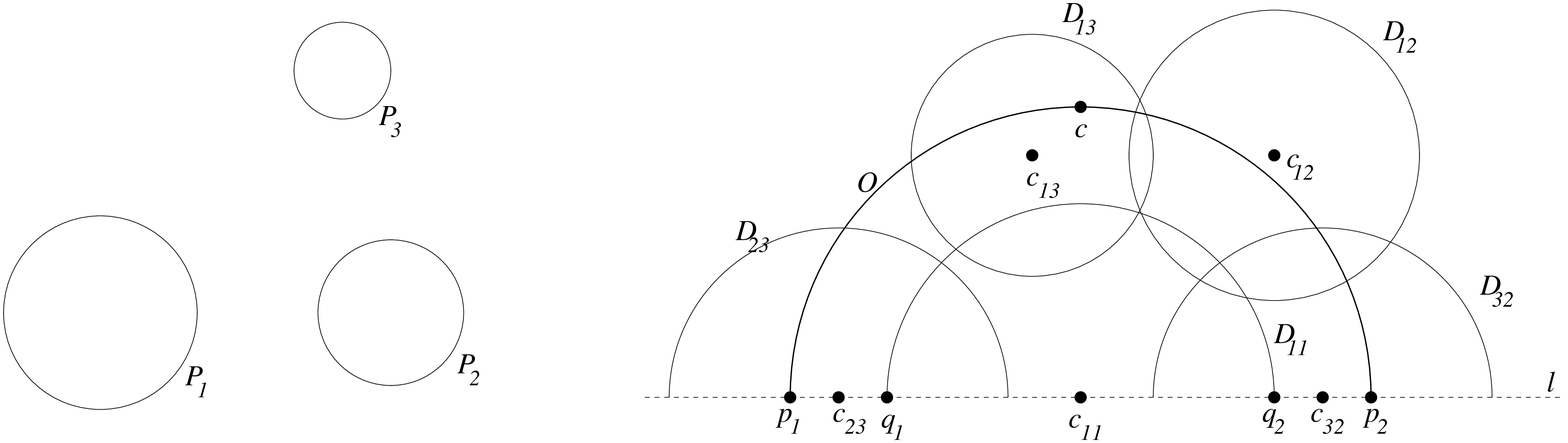}
\caption{{\small A configuration of three pupils and the upper part of its ACS}}
\label{fig:three_pupils}
\end{center}
\end{figure}

\begin{lem}
Among the valid configurations, those in which one radius is half of the objective's radius $R$ and the other two are zero are optimal.
\end{lem}
\begin{proof}
It is straightforward to see that such configurations are
valid. Consider now a configuration in which $\rho_1+\rho_2+\rho_3
< R/2$. We will prove that it cannot be a valid
configuration. Indeed, suppose w.l.o.g. $P_1$ has the largest radius
among three pupils. Then $D_{11}$ is the largest disk among $D_{11},
D_{22}$ and $D_{33}$ and its radius $2\rho_1$ is smaller than $R$. Let
$p_1, q_1, q_2, p_2$ be the intersection points from left to right of
$\partial O$ and $\partial D_{11}$ with $l$ (see
Fig. \ref{fig:three_pupils}). If segment $\overline{p_1q_1}$ is
covered by $D_{23}$, then the diameter of $D_{23}$  is
at least the length of $\overline{p_1q_1}$, i.e., $2(\rho_2+\rho_3)
\geq R - 2\rho_1$ which implies $\rho_1+\rho_2+\rho_3 \geq R/2$
(a contradiction). The case where $\overline{p_2q_2}$ is covered by
$D_{23}$ is symmetrical.  We can therefore assume that $D_{23}$ does
not cover $\overline{p_1q_1}$ nor $\overline{p_2q_2}$, and, by
symmetry, the same holds for $D_{32}$.  Without loss of generality, we
can assume that $D_{12}$ contains $p_1$ or $q_1$ and that $D_{13}$
contains $p_2$ or $q_2$. We denote by $c$ the midpoint of the arc
$p_1p_2$ of $\partial O$. The distance of $c$ to $p_1, q_1, p_2, q_2$
is at least $\sqrt{R^2+(2\rho_1)^2} > R$. Then $c$ is not included in
neither $D_{12}$ nor $D_{13}$ whose diameters are smaller than $R$. It
is not included in $D_{23}$ and $D_{32}$ either since the distance
from $c$ to $c_{23}$ and $c_{32}$ is at least $R$ and the radii of
$D_{23}$ and $D_{32}$ are less than $R$. Hence, the configuration is
not valid.
\end{proof}

It is interesting to see from the above lemma that configurations of three pupils consisting of a pupil of radius $R/2$ and two points are optimal, whatever the position of the pupils may be.

%%%%%%%%%%%%%%%%%%%%%%%%%%%%%%%%%%%%%%%%%%%%%%%%%%%%%%%
\section{An $8\sqrt{2}-$approximation to the smallest number of the pupils of the same radius}
\label{sec:approximation_number_pupils}
In this section, we restrict to the case $\rho_1=\ldots=\rho_n=\rho/2$, then the disks $D_{ij}$ have the same radius $\rho$. We want to find an upper bound for $n$ to cover an objective of radius $R$. As the number of disks is $n^2$, a lower bound $\lceil R/\rho\rceil$ is easily obtained. 

Let $p$ be any prime number, we start by stating a basic property of $p$
\begin{fac}
\label{fac:prime_property}
Let $k,l\in\mathbb{Z}$ such that $\gcd(p,k) = 1$, there exists an integer $0\leq i<p$ satisfying $ik \equiv l \pmod p$.
\end{fac}

\begin{thm}
\label{thm:covering_set}$\{x_i-x_j\mid i,j=0,\ldots,4p-1\}\supseteq\{x\in\mathbb{Z}, |x| < p^2\}$
where
\begin{eqnarray*}
x_k &=& kp + (\frac{k(k + 1)}{2} \bmod p)\\
x_{k+2p} &=& x_k + p,
\end{eqnarray*}
for $k=0\ldots,2p-1$.
\end{thm}
\begin{proof}
Let $x$ be an arbitrary integer between 0 and $p^2-1$, then $x$ can be written as $kp + l$ for some $0\leq k,l < p$. Let $X_i=x_{k+i}-x_{i}$ for $i=0,\ldots,p-1$, we observe that
\begin{eqnarray}
\label{exp:difference_of_centers}
(k - 1)p < X_i < (k + 1)p.
\end{eqnarray}
\begin{displaymath}
X_i \equiv X_0 + ik \pmod p
\end{displaymath}
By Fact \ref{fac:prime_property} there exists some $0\leq i<p$ such that $X_i\equiv l\pmod p$. Hence together with (\ref{exp:difference_of_centers}) the difference of either $x_{k+i}$ or $x_{k+i+2p}$ with $x_i$ will be $x$. The only case where Fact \ref{fac:prime_property} does not apply is when $k = 0$. In this case choose $k = 1$ instead and easily see that the set $\{x_{i+1}-x_{i+2p}\}\cup\{x_{i+1+2p}-x_{i+2p}\}$ generates all integers $1,\ldots,p-1$ and hence contains $x$.
\end{proof}

The above set should not be confused with Golomb ruler \cite{D02} and the set defined by Erd\"os and Tur\'an \cite{ET41} since in the latter sets, the differences between any pair of distinct elements must be unique but do not generally cover all points $1,\ldots,p^2$.

Suppose, w.l.o.g., radius of the disks $\rho = \frac{1}{\sqrt{2}}$ and $R = p^2$ for some prime $p$. Let $\mathcal{S}=\{x\in\mathbb{Z}^2\mid \|x\|_\infty < p^2\}$. We see that the disks of radius $\frac{1}{\sqrt{2}}$ whose centers cover $\mathcal{S}$ are sufficient to cover completely the objective. In other words, we want to find $n$ centers of pupils $c_i\in\mathbb{Z}^2$ such that 
\begin{displaymath}
\{c_i-c_j\mid 1\leq i,j\leq n\} \supseteq \mathcal{S}
\end{displaymath}

\begin{cor}
\label{cor:upper_bound}
$\lceil 8\sqrt{2}R/\rho\rceil$ pupils of radius $\rho$ are sufficient to cover an objective of radius $R$.
\end{cor}
\begin{proof}
The set of pupils is constructed as follows: $c_i=(x_{\lfloor\frac{i}{4p}\rfloor}, x_{i\bmod 4p})$ for $i = 0,\ldots, 16p^2-1$. By applying Theorem \ref{thm:covering_set} first for $x$-coordinate and then for $y$-coordinate, we see that these $16p^2$ pupils are able to cover any element of $\mathcal{S}$ thus the objective of radius $R$. As $R = p^2$ and $\rho = \frac{1}{\sqrt{2}}$, we yield the upper bound.
\end{proof}

The following is an immediate consequence of Corollary \ref{cor:upper_bound} and the lower bound observed earlier.
\begin{cor}
There is an $8\sqrt{2}-$approximation algorithm to cover the objective of radius with the smallest number of pupils of the  same radius.
\end{cor}

%%%%%%%%%%%%%%%%%%%%%%%%%%%%%%%%%%%%%%%%%%%%%%%%%%%%%%%%%%%%%%%%
\section{The fixed-center problem}
\label{sec:fixed-center_problem}

In sections \ref{subsec:same_alpha} and \ref{subsec:different_alphas}, the centers of the pupils are fixed and we present two heuristic algorithms for optimizing the radii among the valid configurations. Both algorithms are based on the fact that the circle of center $c_{ij}$ and radius $\rho_{ij}+\delta_{ij}(p)$ passes through $p$. Then we provide an approximation algorithm with a given error bound and compare it with the heuristic algorithms. We end up the section with a method to maximize the objective while keeping fixed the radii as well as the positions of the pupils.

% SUBSECTION
\subsection{A simple optimization problem}
\label{subsec:same_alpha}
If we increase each of the radii of the pupils by a real number $\alpha/2$, the radii of the disks $D_{ij}$ then increase by $\alpha$ and $Apo(\mathcal{D})$ remains unchanged. Hence there exists a minimal value of $\alpha$, denoted $\alpha^*$, for which the objective is covered by the union of the new (enlarged) disks.

The following shows that $\alpha^*$ can be computed exactly in $O(N\log N)$ time (see Algorithm \ref{alg:same_alpha}). We recall that $V_{ij}$ is the set of vertices of $A_{ij}$ inside $O$ and the intersection points of $\partial A_{ij}$ with $\partial O$. 

\begin{lem}
\label{lem:alpha_star}
\begin{eqnarray*}
\alpha^* = \max_{ij}\max_{p\in V_{ij}}\delta_{ij}(p)
\end{eqnarray*}
\end{lem}
\begin{proof}
It is easy to see that $\max_{ij}\max_{p\in V_{ij}}\delta_{ij}(p)$ is the minimal value of $\alpha$ for which $V_{ij}\subseteq D_{ij}$. The result follows from Lemma 4.
\end{proof}

\begin{algorithm}[]
\caption{Compute $\alpha^*$}
\label{alg:same_alpha}
\begin{algorithmic}[1]
\STATE $\alpha^* \gets -\infty$
\STATE compute $Apo(\mathcal{D})$ and $V_{ij}$
\FORALL {cells $A_{ij}$ of $Apo(\mathcal{D})$}
\FORALL {$x\in V_{ij}$ }
\STATE $\alpha^*\gets \max(\alpha^*, \delta_{ij}(x))$
\ENDFOR
\ENDFOR
\RETURN $\alpha^*$
\end{algorithmic}
\end{algorithm}

%\begin{cor}
%\label{cor:decision_problem2}
%$O$ is covered by $\mathcal{D}$ iff $\alpha^*\leq 0$.
%\end{cor}

%\begin{cor}
%Given the centers of the pupils of the same radius, one can compute in $O(N\log N)$ time their smallest radius to cover a given objective.
%\end{cor}

% SUBSECTION
\subsection{Minimizing the sum of the radii of the pupils}
\label{subsec:different_alphas}

We consider now the more difficult problem of optimizing the sum of the radii of the pupils and propose a heuristic solution that turns out to
perform well in practice.

Instead of increasing the radii of the $P_i$ by a same amount as in the previous subsection, we consider them as $n$ variables. 
Algorithm \ref{alg:different_alphas} below proceeds in two main steps. First, we compute minimal quantities, denoted $\alpha_{ij}$, by which the radii of the $D_{ij}$ must be enlarged/reduced so as to satisfy Lemma \ref{lem:decision_problem} (lines 3--9). This step is similar to Algorithm \ref{alg:same_alpha}. Thanks to the fact that the already visited $\alpha_{ij}$ necessarily increase, the initial $V_{ij}$ will be covered upon termination by the disks $D^{'}_{ij}$ (which are $D_{ij}$ augmented by $\alpha_{ij}$). The objective is then covered by $\bigcup_{i,j=1}^n D^{'}_{ij}$ according to Corollary \ref{cor:disks_cover_cells}. Finally, we want to  minimize the sum of the radii of the $P^{*}_i$  under the constraint that $\rho^{*}_i+\rho^{*}_j$ must be at least the radius of $D^{'}_{ij}$ (line 10):

\begin{eqnarray*}
\textrm{min} && \sum_{i=1}^n\rho^{*}_i \label{*}\\
\textrm{s.t.} && \rho^{*}_i+\rho^{*}_j\geq(\rho_i+\rho_j)+\alpha_{ij},\hspace{1.5cm}i,j=1,\ldots,n\hspace{1cm}(*)\\
&& \rho^{*}_i \geq 0, \hspace{4.6cm} i=1,\ldots,n.
\end{eqnarray*}
Here, $\rho_i$ are the radii of the initial pupils $P_i$ and hence known. This is a linear program whose feasible set is non-empty and bounded. Thus, there exists an optimal solution.

\begin{algorithm}[]
\caption{Minimize the sum of the radii of the pupils}
\label{alg:different_alphas}
\begin{algorithmic}[1]
\STATE $\varepsilon \gets$ any small positive constant
\REPEAT
\STATE $\alpha_{ij} \gets -\infty$, \hspace{0.5cm} $i,j=1,\ldots,n$
\STATE compute $Apo(\mathcal{D})$ and $V_{ij}$
\FORALL {cells $A_{ij}$ of $Apo(\mathcal{D})$}
\FORALL {$x\in V_{ij}$}
\STATE $\alpha_{ij}\gets \max(\alpha_{ij}, \delta_{ij}(x))$
\ENDFOR
\ENDFOR
\STATE compute $\{\rho^{*}_i\}_{i=1,\ldots,n}$ by solving the linear program (*)
\STATE $err \gets \sum_{i=1}^n \rho_i - \sum_{i=1}^n\rho^{*}_i$
\STATE $\rho_i \gets \rho^{*}_i, i = 1,\ldots,n$
\UNTIL $err < \varepsilon$ except for the first iteration
\RETURN $\{\rho^{*}_i\}_{i=1,\ldots,n}$ 
\end{algorithmic}
\end{algorithm}

Note that we need to update the Apollonius diagram since the pupils' radii change after each iteration of the {\bf repeat} loop. 
\begin{lem}
Algorithm \ref{alg:different_alphas} always terminates.
\end{lem}
\begin{proof}
The initial $V_{ij}$ is included in $D^{'}_{ij}$ by the construction of $\alpha_{ij}$. According to Corollary \ref{cor:disks_cover_cells}, $O$ is therefore covered by $\bigcup_{ij}D_{ij}$ after the first iteration. Hence, we may assume that the objective is covered. In this case, Lemma \ref{lem:alpha_star} implies that no $\alpha_{ij}$ is positive which shows that, at each step, $\rho^{*}_i+\rho^{*}_j \leq \rho_i+\rho_j$ and hence $\sum_{i=1}^n\rho^{*}_i\leq\sum_{i=1}^n\rho_i$. Since $\sum_{i=1}^n\rho^{*}_i$ is positive, Algorithm \ref{alg:different_alphas} necessarily terminates after a finite number of iterations.
\end{proof}

\noindent\\{\bf Minimizing the total area of the pupils:} Replacing the objective function $\sum_{i=1}^n\rho^{*}_i$ in (*) with $\pi\sum_{i=1}^n\rho^{*2}_i$ yields a quadratic program which minimizes the total area of the pupils.

\noindent\\{\bf Additional constraints:} In addition to covering the objective, we can also bound the radii of the pupils and forbid any overlap among the pupils. This can be done by adding the following constraints to the linear program (*)

$\rho^{*}_i+\rho^{*}_j\leq\|c_i-c_j\|, \hfill 1\leq i<j\leq n,$

$min\_radius \leq\rho_i\leq max\_radius, \hfill i=1,\ldots,n$.

\noindent Algorithm \ref{alg:different_alphas} has been implemented and appears to work well in practice. Fig. \ref{fig:fixed-center_iterations} compares the results of Algorithms \ref{alg:same_alpha}, \ref{alg:different_alphas} with the optimal solution computed by the following exhaustive search method.

%\subsection{Exhaustive search algorithm on discrete radii}
%\label{sec:exhaustive_search}
\noindent\\{\bf Exhaustive search algorithm:}
If the radii of the pupils are assumed to be integer multiples of a small number $\theta$, then the exhaustive search methods can be applied and the optimal solution in the continuous case must be at least the solution found by these methods minus $n\theta$. We hence have an approximation algorithm within a given error bound.

\subsection{Maximizing the objective}
Now we keep the pupils fixed (radii and positions) and maximize the radius of the objective under the constraint that it is covered by the union of the disks.
\begin{pro}
\label{pro:intersection_point}
If an edge $pq$ of $A_{ij}$ cuts $\partial D_{ij}$ at a point $x\neq p$ and $q$, then there is a point $x'$ on $pq$ that is close to $x$ and not contained in $\mathcal{D}$.
\end{pro}
\begin{proof}
From the fact that $\delta_{ij}(.)$ is a unimodal function and $\delta_{ij}(x) = \delta_\mathcal{D}(x) = 0$.
\end{proof}

The following corollary, whose proof is referred to the full version of the paper, computes the maximal radius $R^*$ of the objective for which it is covered by $\mathcal{D}$. 
\begin{cor}
If $D_{ii}\subseteq A_{ii}$ for some $i=1,\ldots,n$ then $R^* = 2\rho_i$. Otherwise,
\begin{displaymath}
R^* = \min_{ij} \min_{x\in \partial A_{ij}\cap \partial D_{ij}}\| x \|
\end{displaymath}
\end{cor}
%\begin{proof}
%By Proposition \ref{pro:intersection_point} the objectives of radius greater than $R^*$ are not covered.
%To see why the union of the disks covers the objective of radius $R^*$, just construct $\mathcal{D'} = \cup_{i=1}^n D_{ii}$ and then add to $\mathcal{D'}$ the other disks in $\mathcal{D}$. We see that that the distance to the origin from the points on the arc $D_{ij}\cap\partial(\mathcal{D'}\cap D_{ij})$ are greater than from the points of $\|\partial A_{ij}\cap\partial D_{ij}\|$ which is at least $R^*.
%\end{proof}

%%%%%%%%%%%%%%%%%%%%%%%%%%%%%%%%%%%%%%%%%%%%
\section{The fixed-radius problem}
\label{sec:fixed-radius_problem}

\begin{figure}[]
\begin{center}
a)\includegraphics[height=1.5cm]{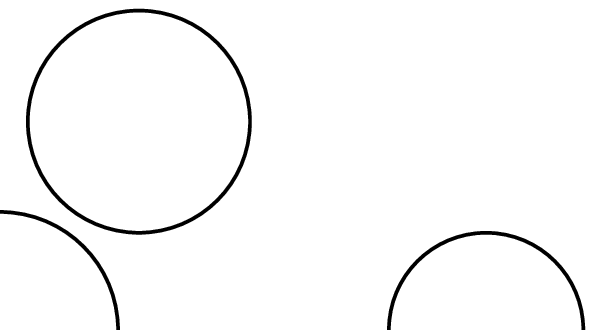}\includegraphics[height=3.2cm]{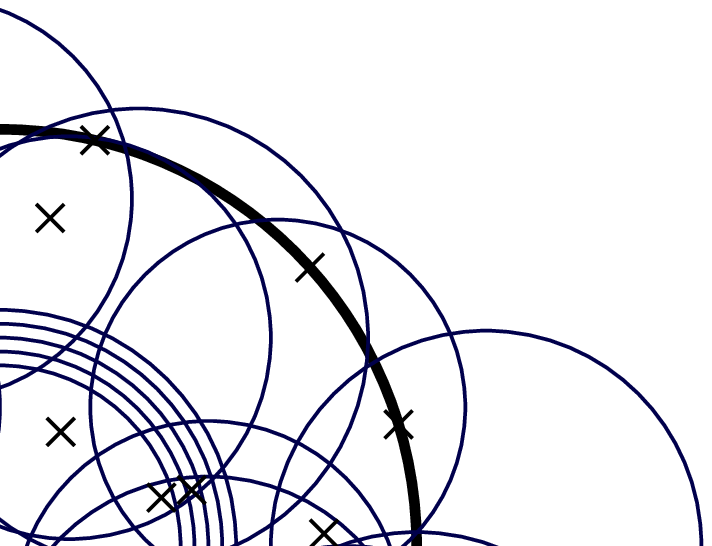}\hspace{1cm}
b)\includegraphics[height=1.435cm]{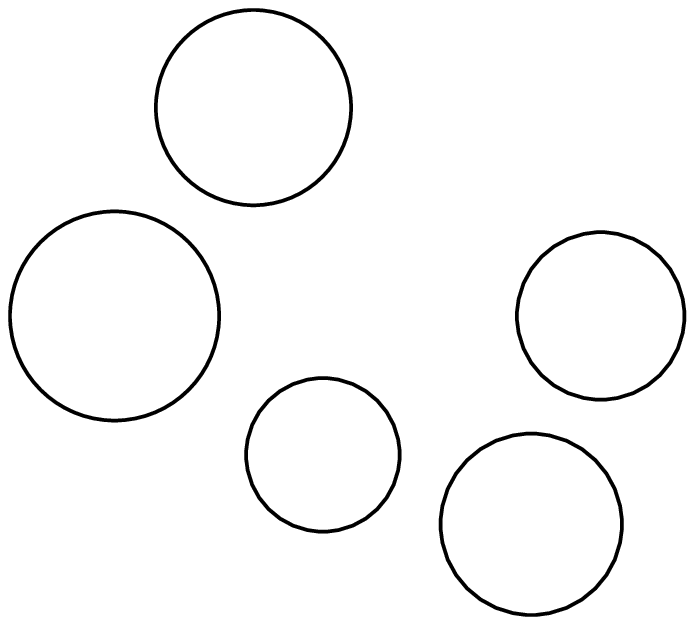}\includegraphics[height=3.06cm]{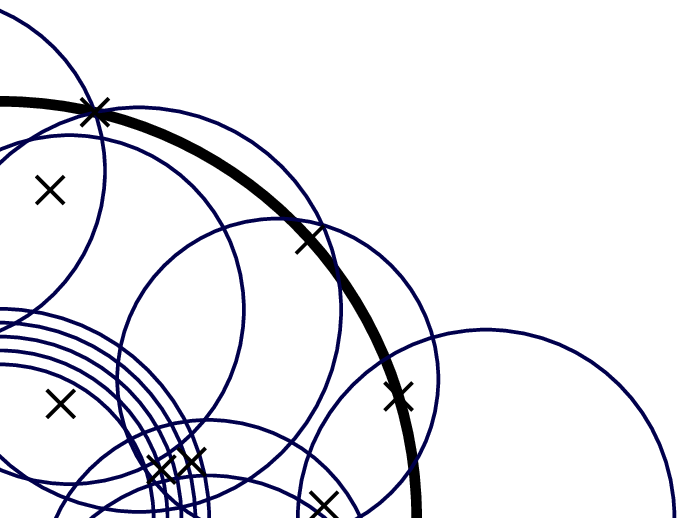}\\
c)\includegraphics[height=1.26cm]{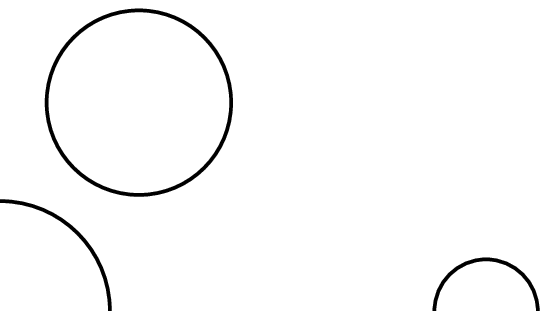}\includegraphics[height=2.69cm]{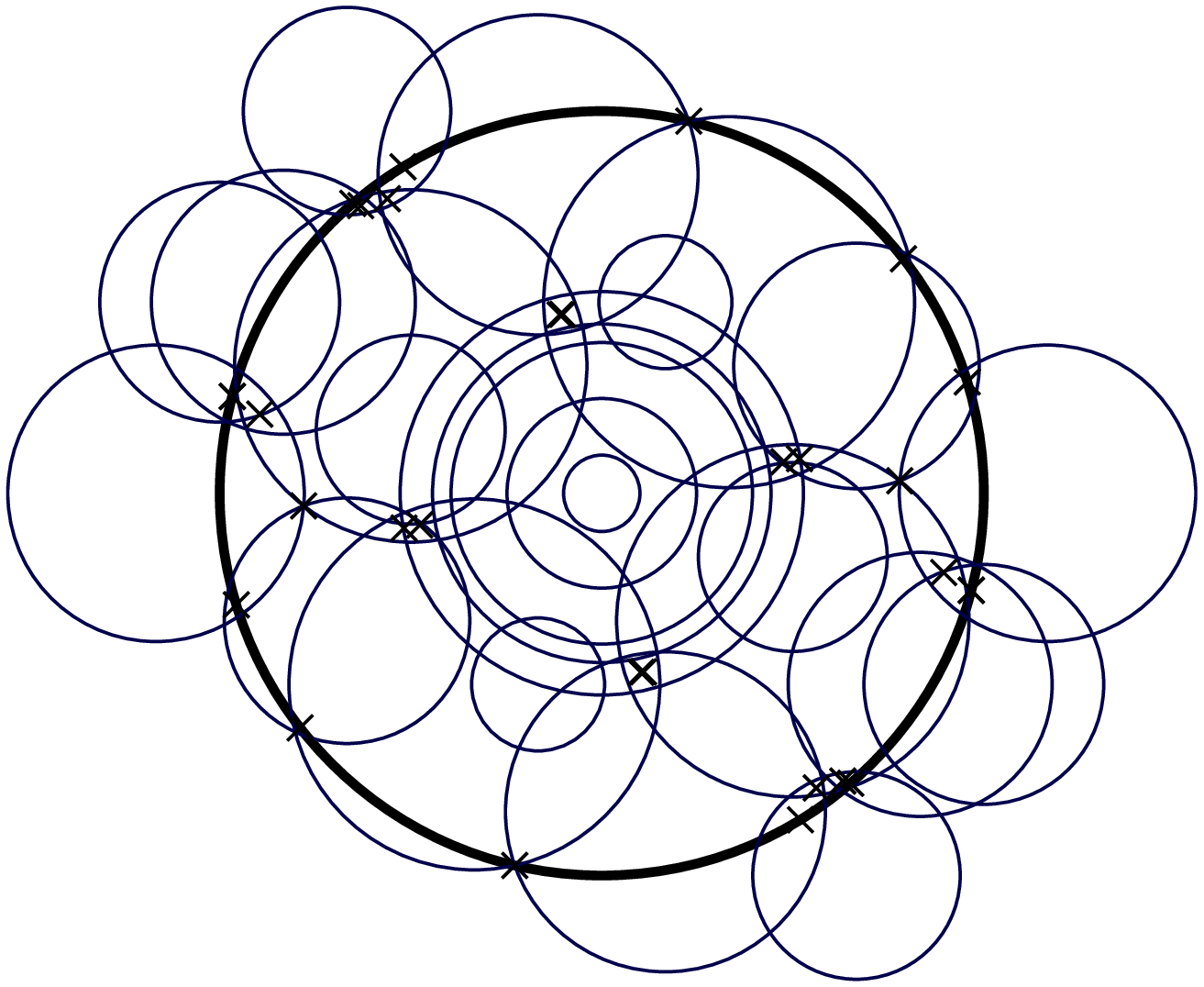}
d)\includegraphics[height=1.295cm]{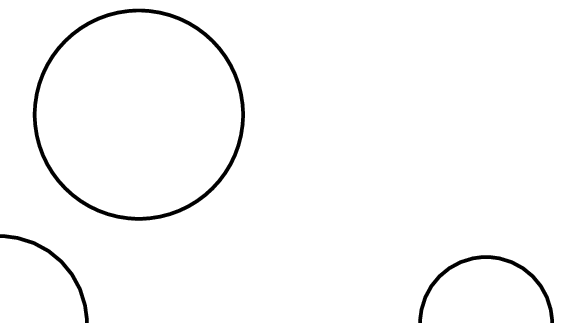}\includegraphics[height=2.76cm]{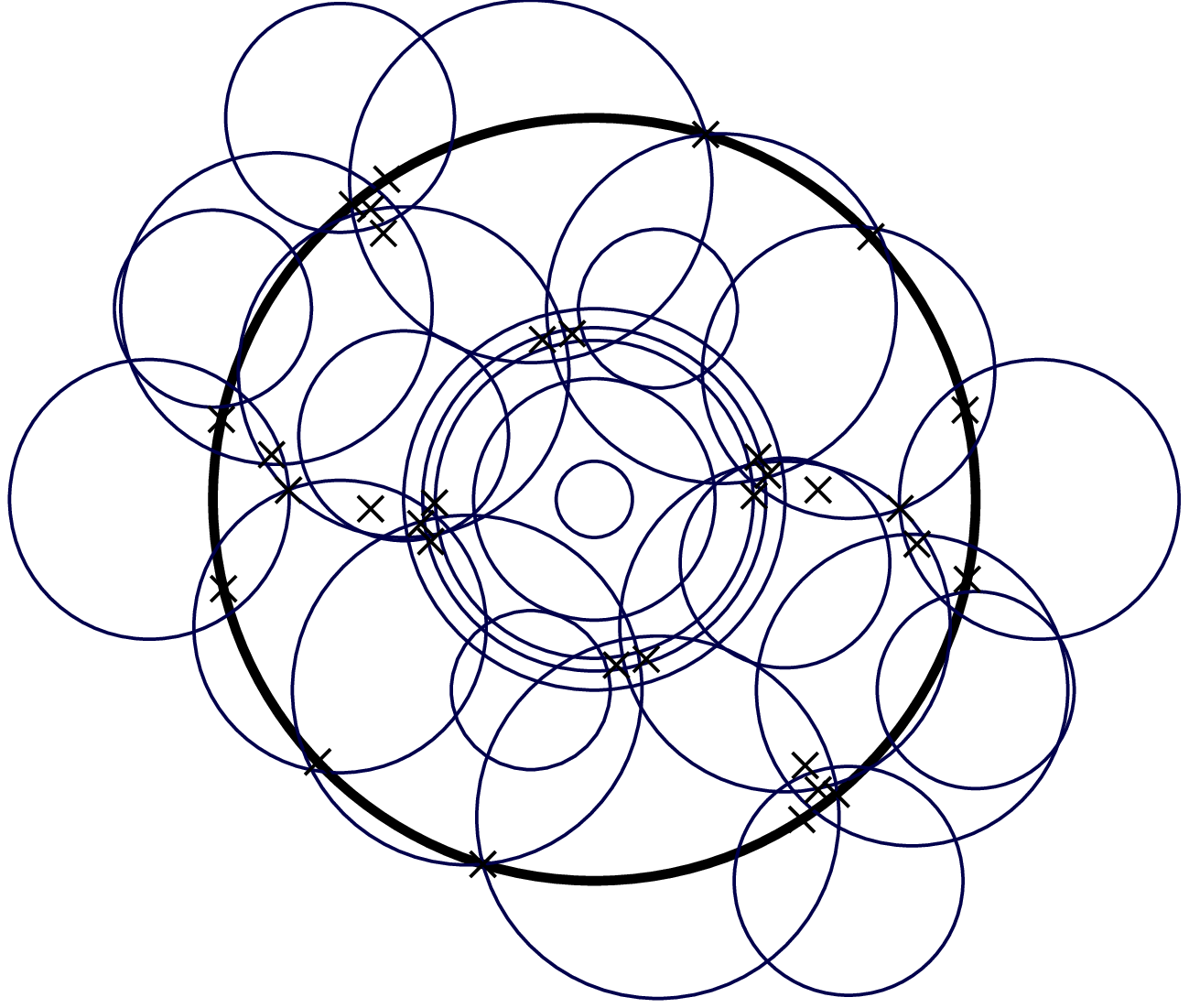}\hspace{0.1cm}
e)\includegraphics[height=1.23cm]{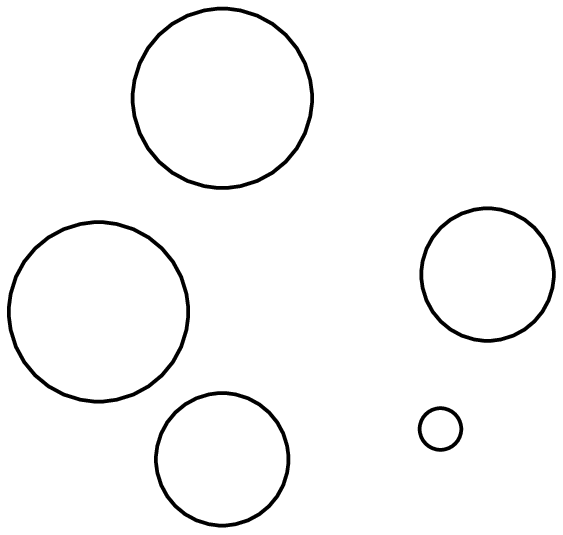}\includegraphics[height=2.62cm]{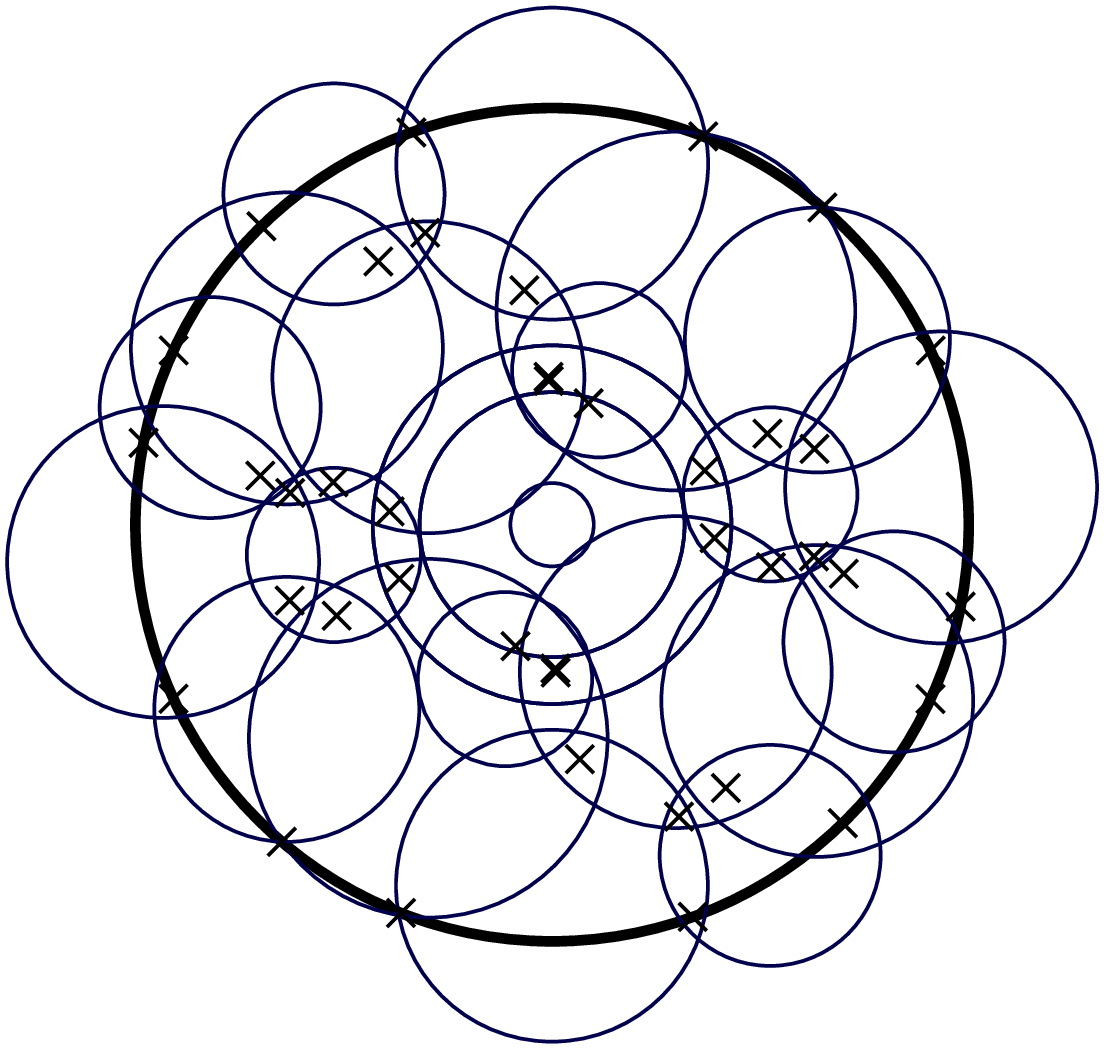}

\caption{{\small {Initial configuration of 5 pupils (a). Results after applying Algorithm \ref{alg:same_alpha} (b), Algorithm \ref{alg:different_alphas} (c) and the exhaustive search algorithm (d). The total areas of the pupils in (a), (b), (c) are 35.6571, 27.1062 and 19.8421 respectively. The optimal solution must be at least $19.572$ as computed by the exhaustive algorithm. The area of pupils in (e) is only 16.4793 when we move the pupils by the algorithm in section \ref{sec:fixed-radius_problem} and then apply Algorithm \ref{alg:different_alphas}.}}}
\label{fig:fixed-center_iterations}
\end{center}
\end{figure}

\begin{figure}[]
\begin{center}
a)\includegraphics[height=1.065cm]{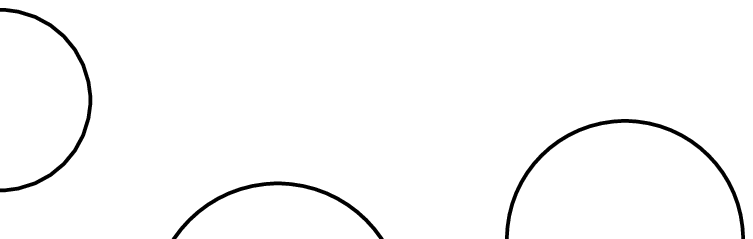}\includegraphics[height=2.93cm]{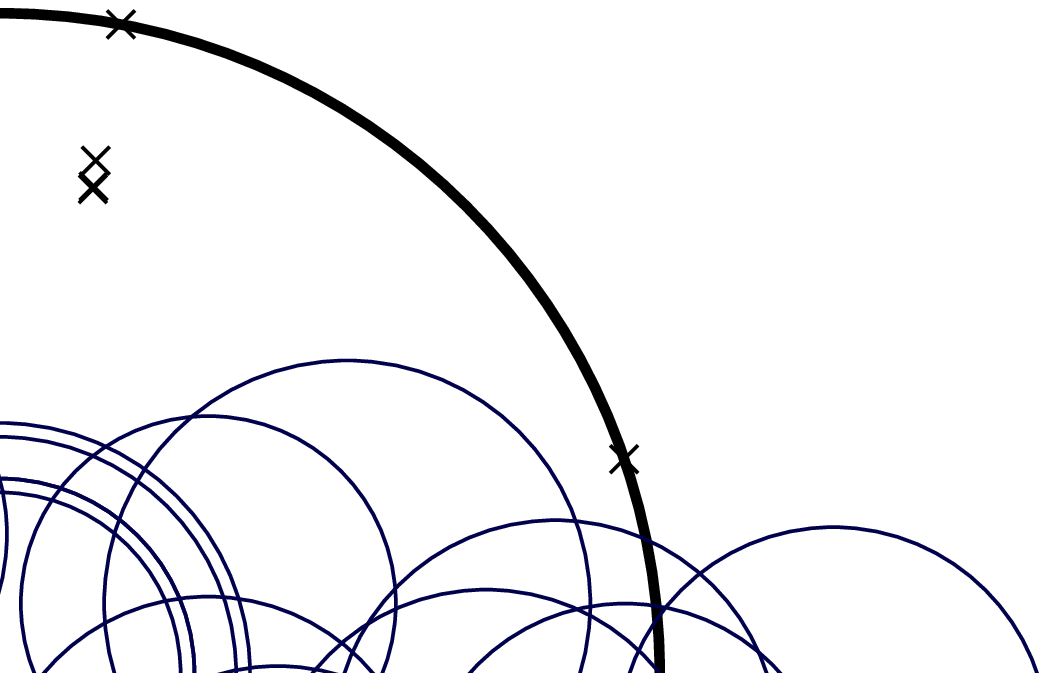}\hspace{0.4cm}
b)\includegraphics[height=1.75cm]{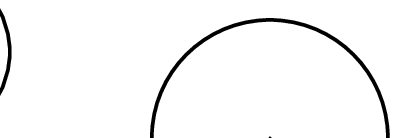}\includegraphics[height=3cm]{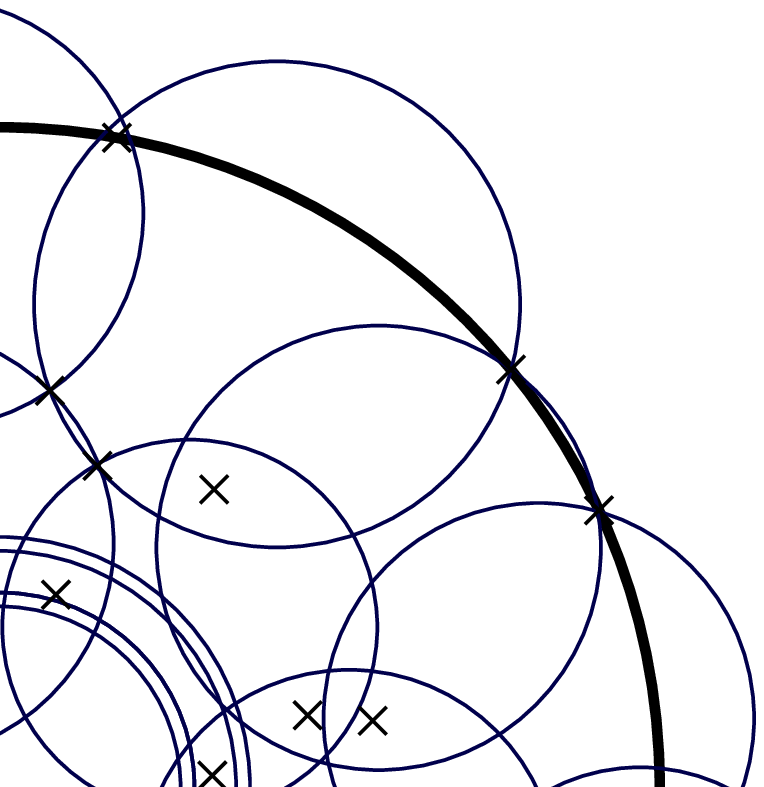}\\
c)\includegraphics[height=1.54cm]{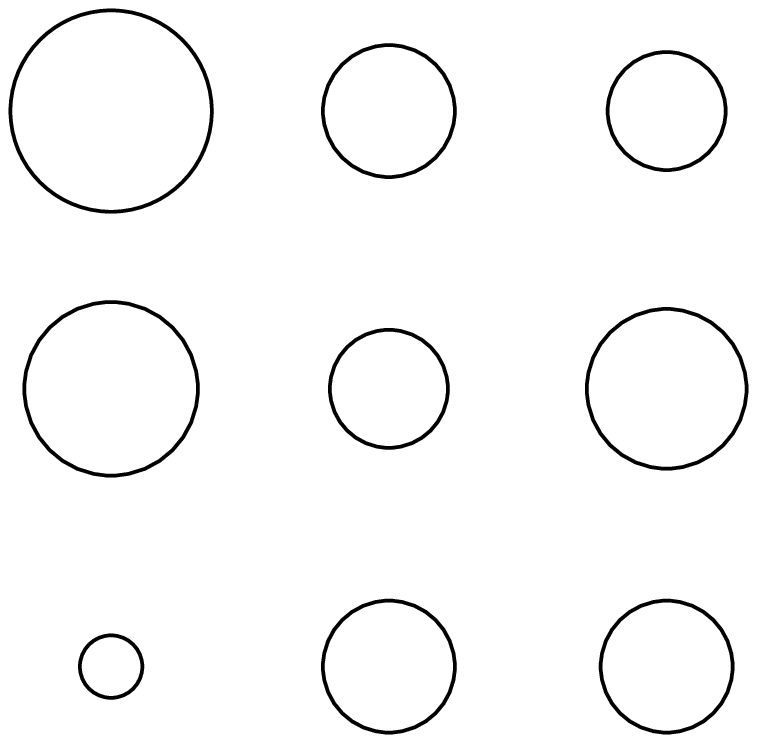}\hspace{0.5cm}\includegraphics[height=3.08cm]{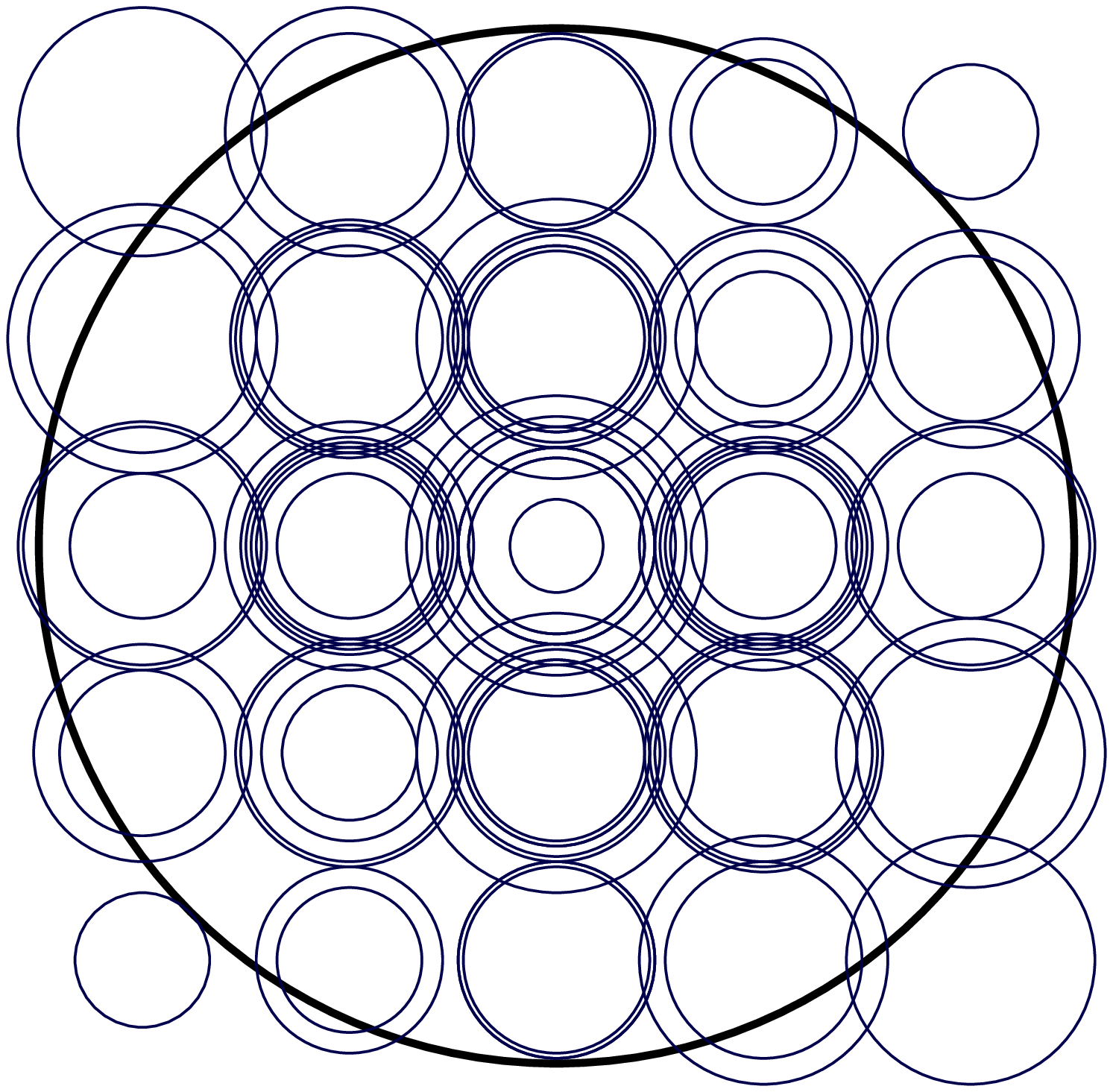}\hspace{0cm}\hspace{2cm}
d)\includegraphics[height=1.75cm]{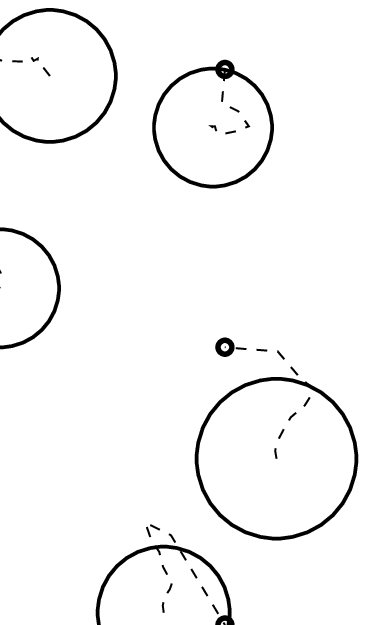}\hspace{0.5cm}\includegraphics[height=3.5cm]{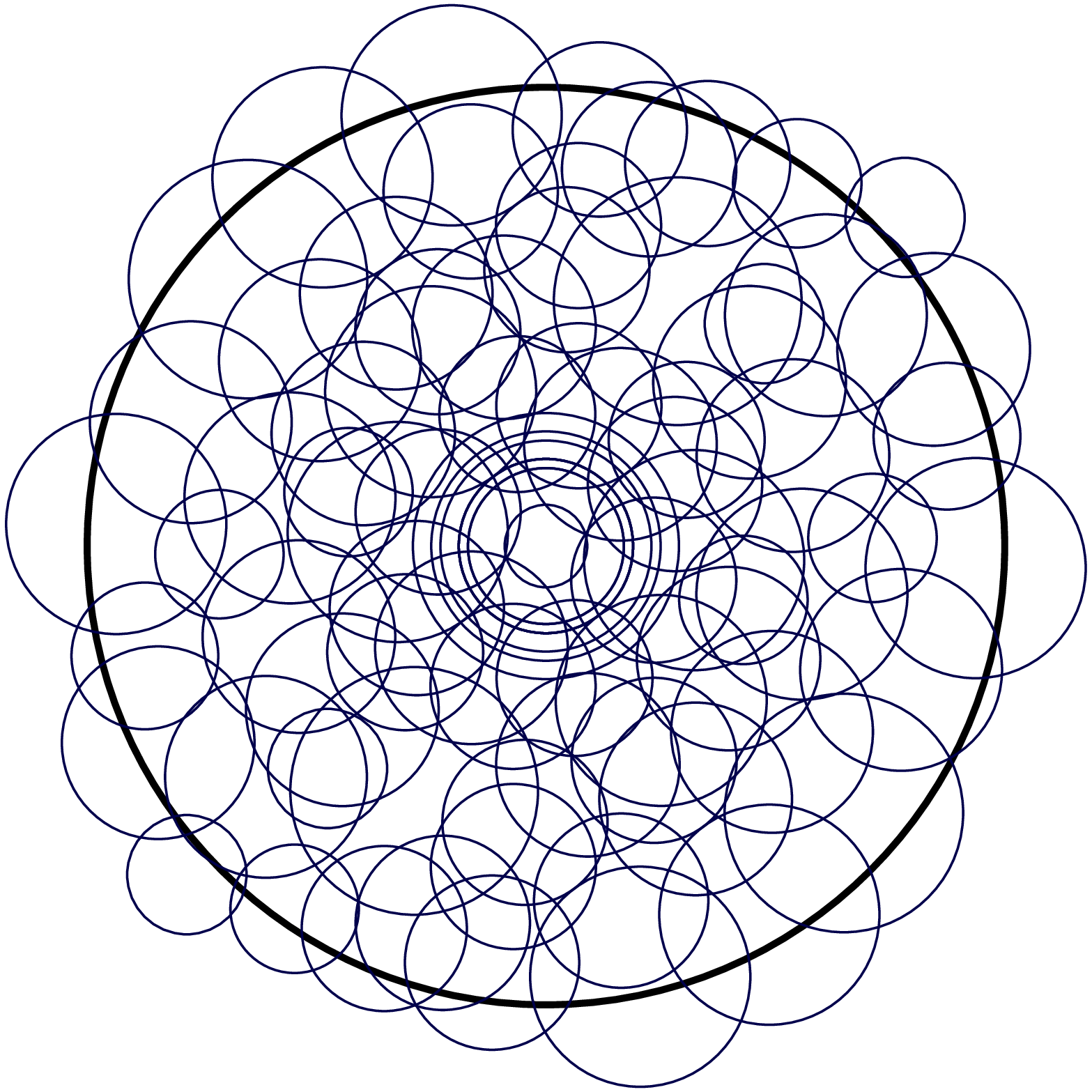}
\caption{{\small {a) The pupil centers are initially placed about a planar line. b) Dotted curves illustrate the movements of the pupils after iterating 24 times the algorithm in section \ref{sec:fixed-radius_problem} when the union of disks cover completely the objective. c) A configuration of 9 pupils d) Result obtained by iterating 9 times the algorithm.}}}
\label{fig:fixed-radius_iterations}
\end{center}
\end{figure}

\begin{figure}[]
\begin{center}
a)\includegraphics[height=1.6cm]{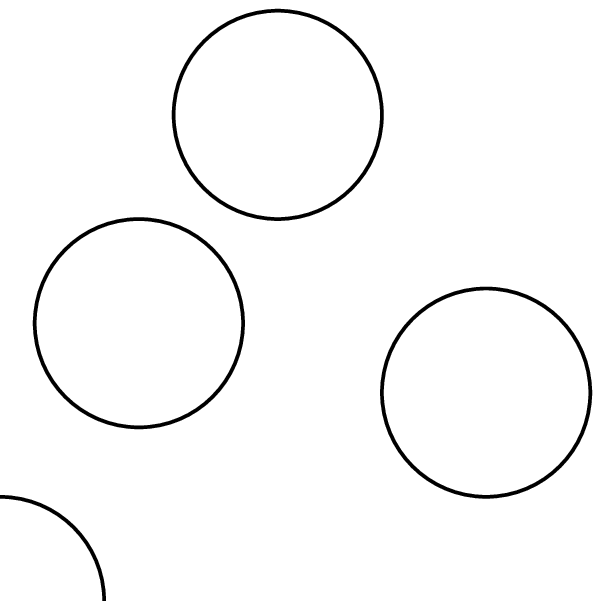}\includegraphics[height=3.2cm]{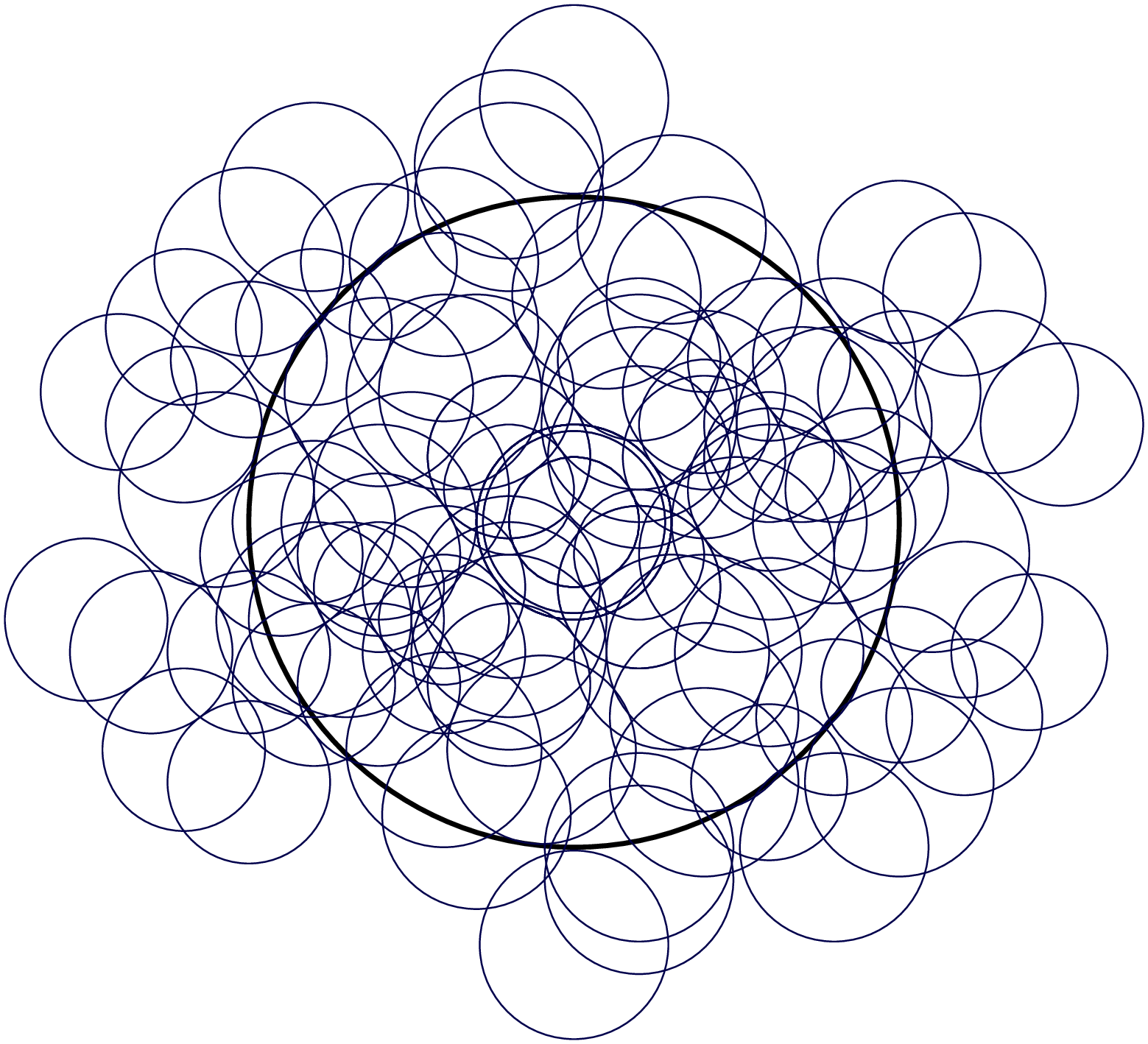}
b)\includegraphics[height=1.59cm]{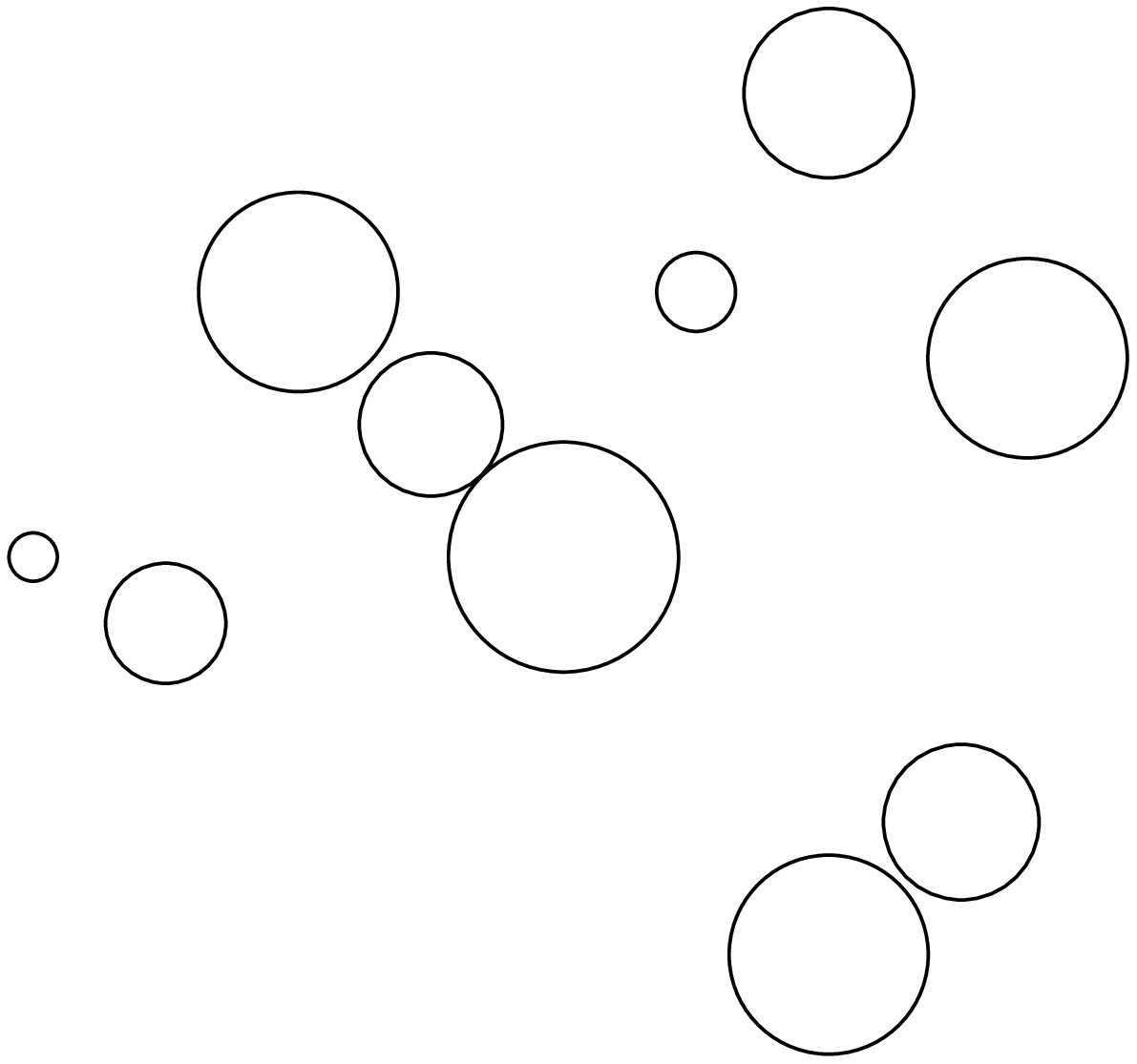}\includegraphics[height=3.18cm]{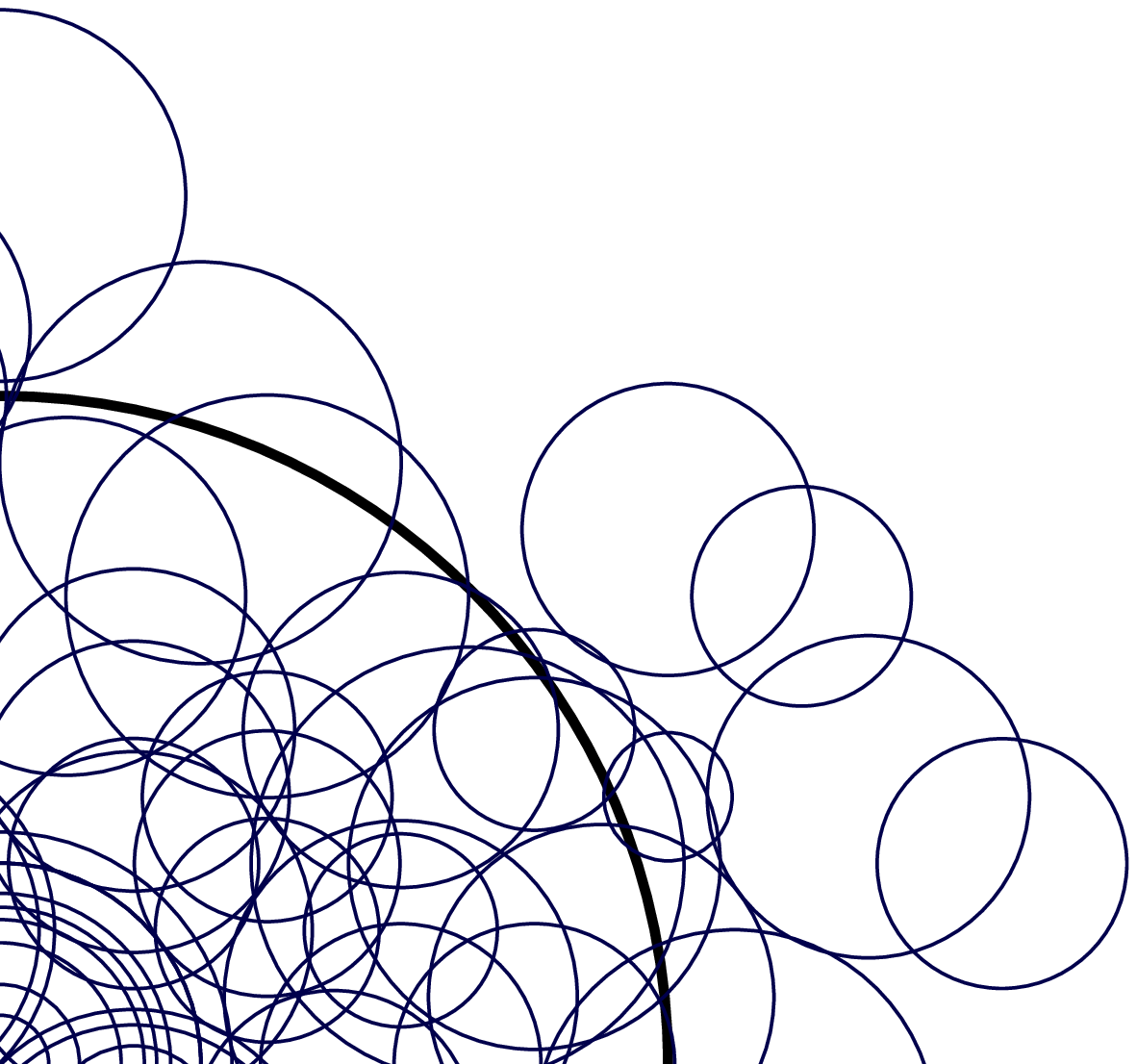}
c)\includegraphics[height=1.4cm]{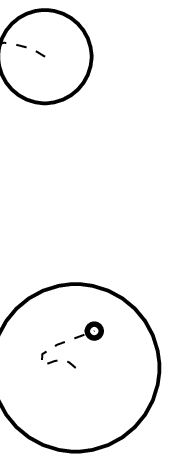}\includegraphics[height=2.80cm]{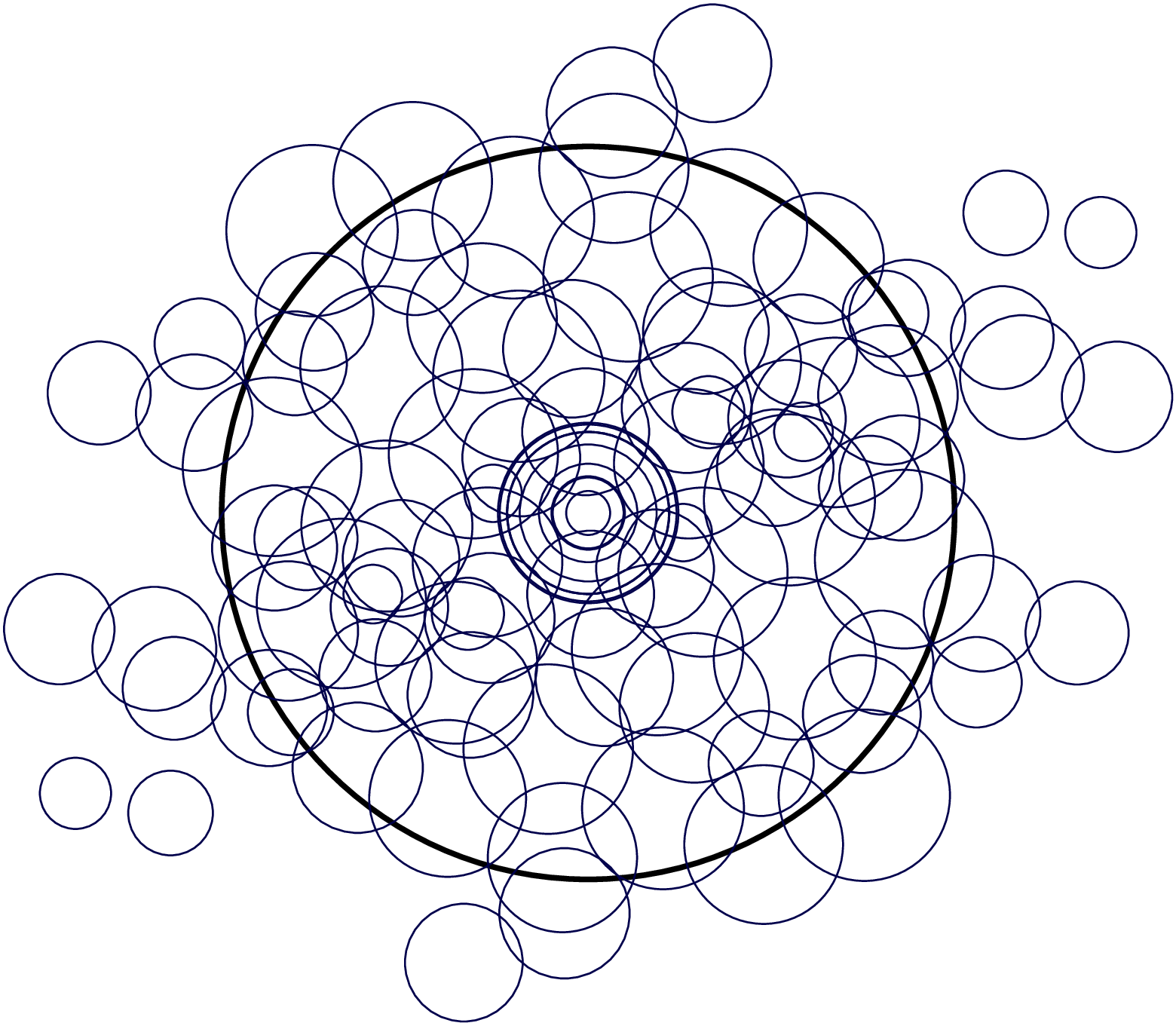}
\caption{{\small {a) A configuration of 10 pupils. b) Result after applying Algorithm \ref{alg:different_alphas}. c) Result after moving the pupils by the algorithm in section \ref{sec:fixed-radius_problem} and then applying Algorithm \ref{alg:different_alphas}. The total areas of the pupils in (a), (b) and (c) are 57.0827, 48.0171 and 26.8088 respectively.}}}
\label{fig:combine_algorithms}
\end{center}
\end{figure}

In this section we fix the radii and propose a heuristic algorithm for moving the set of pupils so that its ACS covers the objective. Our algorithm is based on Corollary \ref{cor:disks_cover_cells}. More precisely, we want to capture the point sets $V_{ij}$ by the disks $D_{ij}$. Given a set of points $P$ and a disk, the optimal center position for the disk to cover $P$ is the point that minimizes the maximal distance to any point of $P$
\begin{equation}
\label{eqn:smallest_enclosing_disk}
\min_{p\in P} \max \|x - p\|.
\end{equation}
This is the so-called smallest enclosing disk problem and a linear algorithm to compute exactly the disk center can be found in \cite{BKOS00}. Unfortunately, function (\ref{eqn:smallest_enclosing_disk}) being non-differentiable makes it hard to apply to our problem. Another approach is to minimize the sum of the squared distance from the center to each point of $P$
\begin{displaymath}
\min{\sum_{p\in P}}\|x - p\|^2.
\end{displaymath}

This function is convex and attains its minimum at the barycenter of $P$. Our algorithm works as follows. We begin with a given configuration of pupils, compute the set $V_{ij}$ and move the pupils to minimize the following function
\begin{equation*}
%\label{eq:optimization_program}
\textrm{min}  \sum_{i,j=1}^n\sum_{p\in V_{ij}}\|(c^{*}_i-c^{*}_j)-p\|^2
\end{equation*}
Here the centers $c^{*}_i$ of the pupils are variables and we recall that $c^{*}_i-c^{*}_j$ becomes the center of disk $D^{*}_{ij}$. The objective function being the sum of convex functions,  is thus convex. We can update the sets $V_{ij}$ and iterate the algorithm until we obtain the desired result. As shown in Fig. \ref{fig:fixed-radius_iterations}, the initial configuration is not critical. The algorithm can also be used as a preprocessing step to improve Algorithm \ref{alg:different_alphas} (see Figs. \ref{fig:fixed-center_iterations}e and \ref{fig:combine_algorithms}).

\vspace{0.1cm}\noindent{\bf ACKNOWLEDGMENT.} 
%\section*{Acknowledgment}
We thank Helmut Alt, G\"unter Rote and Mariette Yvinec for helpful discussions and careful proofreading of early drafts of this paper.

\bibliographystyle{plain}
\bibliography{bibfile}

\end{document}